\documentclass[12pt]{article}
\usepackage{amsmath,amssymb,amsthm}
\ifx\pdfoutput\undefined
\usepackage[dvips]{graphicx}
\else
\usepackage[pdftex]{graphicx}
\fi

\def\C{\mathbb{C}}

\def\Z{\mathbb{Z}}
\def\Q{\mathbb{Q}}
\def\N{\mathbb{N}}
\def\G{{\cal G}}
\def\Id{\hbox{1\kern-2.9pt I}}

\newtheorem{thm}{Theorem}
\newtheorem{lem}[thm]{Lemma}
\newtheorem{cor}[thm]{Corollary}
\theoremstyle{definition}

\begin{document}
\author{Alexandre Stefanov}
\title{Special Symplectic Subgroup over Integers Arising as a Factor of
The Braid Group}
\date{October 18, 2004}
\maketitle

\section{Introduction}

The braid group can be defined in many ways and since its introduction
in~\cite{Art1,Art2}, one of them is a representation in the group of
automorphisms of the free group. If the generators of the free group
are $g_1,g_2,\dotsc,g_{n+1}$, the action of the braid group is given by
\begin{equation}
\label{braid}
\begin{array}{rcl}
\sigma_i(g_i) &=&g_ig_{i+1}g_i^{-1}\\
\sigma_i(g_{i+1}) &=&g_i\\
\sigma_i(g_j) &=&g_j,\quad |i-j|\ge 2.\\
\end{array}
\end{equation}
Here $\sigma_1$, \dots, $\sigma_n$ are the standard generators of the braid
group $B_{n+1}$ on $n+1$ strands.

This action can be considered also over any ordered set of elements of a 
group. It appears when one considers a fiber bundle $\pi:M\to\C\setminus
\{p_1,\dotsc,p_{n+1}\}$ over the plane with $n+1$ punctures. Let the bundle
be equipped with a complete flat connection and $M_t=\pi^{-1}(t)$ to be
the fiber of $M$ over $t$. For any loop $\gamma$ based at $t$, we may
integrate the connection on $M$ along $\gamma$ to obtain a homomorphism
\begin{equation}
a_\gamma:M_t\to M_t
\end{equation}
which, as the connection is flat, depends only on the homotopy class of
$\gamma$. In this way a representation of the fundamental group is obtained
\begin{equation}
\pi_1(C\setminus\{p_1,\dotsc,p_{n+1}\})\to Aut(M_t)
\end{equation}
called the monodromy representation. The braid group enters when one
considers continuous deformation of the points $p_1,\dotsc,p_{n+1}$ and the
connection, in a way preserving the monodromy. The possibility of such
deformations for systems of Fuchsian differential equations is proved
in~\cite{Sch} by constructing an integrable Pfaffian
differential equation, which the
coefficients must satisfy in order to preserve the monodromy. Solutions to
these equations correspond to orbits of the braid group on tuples of
linear transformations $(A_1,\dotsc,A_{n+1})$, $A_i\in GL(V)$, subject to
the equivalence $(A_1,\dotsc,A_{n+1})\sim (AA_1A^{-1},\dotsc,AA_{n+1}A^{-1})$,
as the monodromy is fixed only up to simultaneous
conjugation. Because of this, the braid
\begin{equation}
\Delta^2=(\sigma_1,\dotsc,\sigma_n)^{n+1}:(A_1,\dotsc,A_{n+1})\mapsto
(AA_1A^{-1},\dotsc,AA_{n+1}A^{-1})
\end{equation}
where $A=r_1\dotsm r_{n+1}$, will act trivially. It is known~\cite{Bir} that
the braid $\Delta^2$ generates the center of $B_{n+1}$.

One class of linear transformations, on which the action of the braid group
is particularly simple, is that of reflections. A reflection in a linear
space $V$, equipped with a nondegenerate symmetric bilinear form, can
be written as $r=\Id-\langle v|$, where the vector $v$ satisfy
$\langle v|v\rangle=2$. The relative position of $n+1$ reflections
$r_1,\dotsc,r_{n+1}$, $r_i=\Id-\langle v_i|$ is be specified by their Gram
matrix
\begin{equation}
G_{ij}={\langle v_i|v_j\rangle}\,.
\end{equation}
We arrive at an action of the braid group, factored over its center, on the
Gram matrices. It has the following form
\begin{equation}\label{sym}
\sigma(G)=K_\sigma(G)\cdot G\cdot K_\sigma(G)\,.
\end{equation}
The symmetric matrices $K_\sigma(G)$ depend on the braid as well as on
the Gram matrix, therefore the action is nonlinear on the entries of the Gram
matrix.

In~\cite{DM} were found all finite orbits of the braid group action on triples
of reflections, having nondegenerate Gram matrix. It was shown that these
orbits correspond to pairs of reciprocal
regular polyhedra or star-polyhedra, while the elements in the orbits
correspond to the Schwarz triangles (see~\cite{Cox}). Motivated by this result,
in the present article we study the
orbits of the braid group on arbitrary number of reflections with Gram
matrix of rank~2. This is the first nontrivial case with respect to the rank
as the orbits on rank~1 matrices are trivial.

In the next section 
is introduced angular parametrization of the rank~2 Gram matrices. It is
shown that the action of the braid group $B_{n+1}$ is linear on these
parameters yielding a representation into the group of integer valued matrices 
with determinant one
\begin{equation}
\rho:B_{n+1}\to SL_n(\Z)\,.
\end{equation}
This can be obtained from the Burau representation~\cite{Bur} by substituting
$t=-1$. It has been considered in~\cite{Arn} for even $n$, while for odd $n$
their representation differs from the one considered here.

Then it is shown that there is an antisymmetric form on the lattice $\Z^n$,
preserved by the action of $\rho(B_{n+1})$. This is a surprising result as
we begin with the braid group action on orthogonal reflections and arrive
at representation in the integer valued symplectic group
$\rho:B_{n+1}\to Sp_n(\Z)$ for even $n$. For odd $n$ the representation is
reducible $\rho:B_{n+1}\to Sp_{n-1}(\Z)\ltimes\Z^{n_1}$.

In the last section are found the finite orbits of the braid group on the
Gram matrices of rank~2 and the linear representations are characterized
by showing that the image $\G_n=\rho(B_{n+1})$ of the braid group is a
congruence subgroup of level~2 i.e. it contains the principal subgroup of
level~2
\begin{equation}
\Gamma_0(2)=\ker(Sp_n(\Z)\to Sp_n(\Z/2\Z))
\end{equation}
where it is meant the natural homomorphism by reducing the entries to modular
arithmetics. For odd $n$, correspondingly, the principal congruence subgroup is
\begin{equation}
\Gamma_0(2)=\ker(Sp_{n-1}(\Z)\ltimes\Z^{n_1}\to Sp_{n-1}(\Z/2\Z)\ltimes
(\Z/2\Z)^{n_1})
\end{equation}
However it is not modular reductive i.e. the matrices in $\G_n$ do not have
the form 
\begin{equation}
M=\begin{pmatrix} A & *\\ 0 & B\end{pmatrix}\mod 2
\end{equation}
in any basis, which is the best known example of congruence
subgroups.

\section{Linearization of the action of the braid group}

Let there is a collection of $n+1$ reflections, preserving a nondegenerate
(not necessarily positive definite) symmetric bilinear form $\langle\cdot|
\cdot\rangle$. In such case the reflections must have the form
\begin{equation}\label{refl}
r_i=\Id-\langle v_i|
\end{equation}
for some vectors $v_1,\dotsc,v_{n+1}$, such that $\langle v_i|v_i\rangle=2$.
The relative position of these reflections is given by the Gram matrix
$G_{ij}=\langle v_i|v_j\rangle=2$.
Substituting~\eqref{refl} in~\eqref{braid} and calculating
the Gram matrix of the transformed vectors we obtain
\begin{equation}
\begin{array}{lcl}
\sigma_i(G)_{i\,j}&=&G_{i+1\,j}-G_{i\,i+1}G_{i\,j}\,,\quad j\ne i,i+1\\
\sigma_i(G)_{i+1\,j}&=&G_{ij}\,,\quad j\ne i\\
\sigma_i(G)_{i\,i+1}&=&-G_{i\,i+1}\\
\sigma_i(G)_{j\,j}&=&G_{j\,j}=2\\
\sigma_i(G)_{k\,j}&=&G_{k\,j}\,,\quad k\neq i,i+1\,.
\end{array}
\end{equation}
where the symmetric matrix $K_\sigma$ for the generators of the braid group
is given by
\begin{equation}
(K_{\sigma_i}(G))_{j\,k}=\delta_{j\,k}-\delta_{i\,j}\delta_{j\,k}(1+G_{i\,i+1})
-\delta_{i+1\,j}\delta_{j\,k}+\delta_{i\,j}\delta_{i+1\,k}+
\delta_{i\,k}\delta_{i+1\,j}\,.
\end{equation}
These transformations can be written compactly as~\eqref{sym}.

\begin{thm}
Every symmetric matrix $G$ of rank 2 for which $G_{ii}=2$ can be written
\begin{equation}\label{param}
G_{ij}=2\cos(\phi_i-\phi_j).
\end{equation}
for some complex $\phi_1,\phi_2,\dotsc$
\end{thm}
\begin{proof}
We let $G_{ij}=2g_{ij}$ for ease of notation.
Every $3\times3$ submatrix of $G$ must be degenerate therefore also
every $3\times3$ submatrix of $g$ will be degenerate. Taking the 1-st, $i$-th
and $j$-th row and column
\begin{multline}
\left|\begin{array}{ccc}1&g_{1i}&g_{1j}\\g_{1i}&1&g_{ij}\\g_{1j}&g_{ij}&1
\end{array}\right|=1+2g_{1i}g_{1j}g_{ij}-g_{1i}^2-g_{1j}^2-g_{ij}^2\\
=(g_{1i}^2-1)(g_{1j}^2-1)-(g_{ij}-g_{1i}g_{1j})^2=0\,.
\end{multline}
We write $G_{ij}=2\cos\phi_{ij}$. The above identity implies
\begin{equation}
\cos(\phi_{ij})-\cos(\phi_{1i})\cos(\phi_{1j})=\pm\sin(\phi_{1i})
\sin(\phi_{1j})
\end{equation}
hence
\begin{equation}
\phi_{ij}=\phi_{i}+\epsilon_{ij}\phi_{j}\,\quad\epsilon_{ij}=\epsilon_{ji}
=\pm1\,,
\end{equation}
where $\phi_{i}=\phi_{1i}$. Taking another submatrix of the rows numbered
$1,i,k$ and the columns numbered $1,i,j$
\begin{multline}\label{det2}
\left|\begin{array}{ccc}1&g_{1i}&g_{1j}\\g_{1i}&1&g_{ij}\\g_{1k}&g_{ik}&g_{jk}
\end{array}\right|=g_{1i}g_{ij}g_{1k}+g_{1i}g_{1j}g_{ik}+g_{jk}-
g_{1j}g_{1k}-g_{ij}g_{ik}-g_{1i}^2g_{jk}\\
=-(\epsilon_{jk}+\epsilon_{ij}\epsilon_{ik})\sin^2\phi_i\sin\phi_j\sin\phi_k=0
\,.
\end{multline}

The cosine is an even function so we may take $\epsilon_{1i}=-1$. The sign
of $\epsilon_{ij}$ does not matter if $\phi_i\equiv0\mod\pi$ so we let
$\epsilon_{ij}=-1$ in such case i.e. when $\sin\phi_i\sin\phi_j=0$.
We have $\phi_1\equiv\phi_2\equiv\dotsb\equiv\phi_{k-1}\equiv0\mod\pi$,
$\phi_k\not\equiv0\mod\pi$
for some $k\ge2$. If $\epsilon_{ki}=1$ for some $i$ we change
\begin{equation}
\left\{\begin{array}{rl}\phi_i&\mapsto2\pi-\phi_i\\
\epsilon_{ki}&\mapsto-\epsilon_{ki}=-1\,.\end{array}\right.
\end{equation}

In this way we assure $\epsilon_{ki}=-1$ for every $i$.
From \eqref{det2} it follows that
\begin{equation}
\epsilon_{ij}=-\epsilon_{ki}\epsilon_{kj}=-1\,,
\end{equation}
when $\sin\phi_i\sin\phi_j\ne0$ but we have set $\epsilon_{ij}=-1$ in the
other case too.
\end{proof}

In our problem the matrix $G$ defines the relative position of reflections. 
Each reflection given by a vector $v$ can be given also by $-v$, so there is
an equivalence between matrices $G$ defining the same reflection arrangement
\begin{equation}
G\simeq G'\quad{\rm iff}\quad G'_{ij}=\lambda_i\lambda_jG_{ij}\quad
\lambda_i=\pm1
\end{equation}
In our parameterization this equivalence allows
the change
\begin{equation}\label{lesspi}
\phi_i\mapsto\phi_i-\pi\,,\quad B_{jk}\mapsto\lambda_j\lambda_kB_{jk}\,,\,
\lambda_j=\left\{\begin{array}{rl}-1,&j=i\\1,&j\ne i\,.\end{array}\right.
\end{equation}
Another freedom in the parameterization by angles $\phi_i$ is due to the
appearance of only cosine function of them which is an even function and
allows the simultaneous inversion of their signs.

The action of standard generators of $B_{n+1}$ on the angles $\phi_i$
is given by
\begin{equation}\label{transfi}
\sigma_i:\left\{\begin{array}{lll}
\phi_i & \mapsto & 2\phi_i-\phi_{i+1}\\
\phi_{i+1} & \mapsto & \phi_i\\
\phi_j & \mapsto & \phi_j\quad j\ne i,i+1\,.
\end{array}\right.
\end{equation}
These transformations define a linear representation of $B_{n+1}$
on $\C^{n+1}$. The angles $\phi_i$ parameterize matrices
$B_{ij}=2\cos(\phi_i-\phi_j)$ therefore we have the identification
$\phi_i\equiv\phi_i+2\pi$. Using \eqref{lesspi} we further identify
$\phi_i\equiv\phi_i+\pi$. In this context the action \eqref{transfi} must be
considered over $(\C/\pi\Z)^{n+1}$.

\begin{thm}
The action of $B_{n+1}$ on the parameters $\phi_1,\phi_2,\dotsc,\phi_{n+1}$,
$\phi_i\in\C/\pi\Z$ defined by \eqref{transfi} will have finite orbit if and
only if $\forall i\,,\;\phi_i-\phi_{i+1}\in\pi\Q$.
\end{thm}
\begin{proof}
The matrices of the transformations \eqref{transfi} are unipotent with
a common eigenvector $\phi_1=\phi_2=\dotsb=\phi_{n+1}$ corresponding to
a Gram matrix of rank 1. Apart from this case there exist
some $i$ such that $\phi_i-\phi_{i+1}\ne0$. There must be a power $\sigma_i^k$,
which acts trivially on $(\C/\pi\Z)^{n+1}$ in order to have a finite orbit.
\begin{equation}
\sigma_i^k:\left\{\begin{array}{lll}
\phi_i & \mapsto & (k+1)\phi_i-k\phi_{i+1}\\
\phi_{i+1} & \mapsto & k\phi_i-(k-1)\phi_{i+1}\\
\phi_j & \mapsto & \phi_j\quad j\ne i,i+1\,.
\end{array}\right.
\end{equation}
We have
\begin{equation}
\left|\begin{array}{ccc}
(k+1)\phi_i-k\phi_{i+1}&\equiv&\phi_i\mod\pi\\
k\phi_i-(k-1)\phi_{i+1}&\equiv&\phi_{i+1}\mod\pi\end{array}\right.\quad
\Rightarrow\quad\phi_i-\phi_{i+1}=\frac{p}{k}\pi
\end{equation}
Parameters obeying the above condition on their differences, are written as
$\phi_i=\phi+\frac{p_i}{q_i}\pi$, where $\phi$ is a common phase, preserved by
the transformations~\eqref{transfi}. Let $m=\mbox{lcm}(q_1,\dotsc,q_{n+1})$. We
write $\phi_i=\phi+\frac{r_i}{m}\pi$, and we may always take $0\le r_i<m$
because of the identification $\phi_i\equiv\phi_i+\pi$.
The transformations~\eqref{transfi} preserve $\phi,m$ and as there are finite
number of values for $r_1,\dotsc,r_{n+1}$ the condition of the theorem is also
sufficient.
\end{proof}

Up to here we have not used the fact that in~\eqref{param} enter only the
differences $\phi_i-\phi_j$. Calling $f_i=\phi_{i-1}-\phi_1=\frac{k_i}{m}$
we obtain that the Gram matrices of rank~2 belonging to finite orbits of
the braid group are written as
\begin{equation}
G_{11}=2,G_{1i}=G_{i1}=2\cos\left(\frac{k_{i-1}}{m}\pi\right)\,,\quad
G_{ij}=2\cos\left(\frac{k_{i-1}-k_{j-1}}{m}\pi\right),
\end{equation}
where $k_i\in\Z_m$.

The generators of $B_{n+1}$ transform
the parameters $ k_1, k_2,\dotsc, k_n$ in the following
way

\begin{align}\label{act}
\sigma_1:\left\{\begin{array}{ll} k_1&\mapsto  k_1\\
 k_j&\mapsto  k_j + k_1\,,j>1\end{array}\right.
&\quad
\sigma_i:\left\{\begin{array}{lll} k_{i-1}&\mapsto & 2 k_{i-1}
- k_i\\
 k_i&\mapsto &  k_{i-1}\\
 k_j&\mapsto &  k_j\quad j\ne i,i+1\,,
\end{array}\right.
\end{align}
where $i>1$.

In order to classify the orbits of this image of the braid group on
$\Z_m^{n}$ it is useful first to consider these transformations on
$\Z^{n}$. The action \eqref{act} on $\Z^{n}$ can be interpreted by
considering reflections about points on the real line with integer coordinates
$r_i(x)=2k_i-x\,,\,k_i\in\Z$. Each reflection is determined by the coordinate
$k_i$ of the point it stabilizes. The action \eqref{braid} of the braid group
on these reflections induce transformations on the coordinates of their stable
points, and their differences transform according to~\eqref{act}.

The obtained linear representation of the braid group factored over its
center is not faithful -- there are additional relations between generators
\begin{gather}\label{genrel}
(\sigma_i\sigma_{i+1})^6=(\sigma_i\sigma_{i+1}\sigma_i)^4=\Id\\
(\sigma_i\sigma_{i+1}\dotsm\sigma_{i+2k-1})^{2(2k+1)}=\Id\,,
\end{gather}
which together with the standard relations (\cite{Bir}) between the generators
\begin{equation}\label{genrel1}
\sigma_i\sigma_{i+1}\sigma_i=\sigma_{i+1}\sigma_i\sigma_{i+1}\qquad
\sigma_i\sigma_j=\sigma_j\sigma_i\,,\,j\ne i,i+1
\end{equation}
\begin{equation}
(\sigma_1\sigma_2\dotsm\sigma_n)^{n+1}=\Id
\end{equation}
determine it as an abstract group. Notice that one of the
relations in \eqref{genrel} is redundant as
\begin{equation}
(\sigma_i\sigma_{i+1})^6=(\sigma_i\sigma_{i+1}\sigma_i\sigma_{i+1}
\sigma_i\sigma_{i+1})^2=(\sigma_i\sigma_{i+1}\sigma_i
\sigma_i\sigma_{i+1}\sigma_i)^2=(\sigma_i\sigma_{i+1}\sigma_i)^4
\end{equation}
using \eqref{genrel1}.

Let us call this linear group $\G_n$. Clearly the inverses of $\sigma_i$ are
also represented by matrices with integer entries hence $\G_n\subset GL(n,\Z)$.
We are interested in the orbits of $\G_n$ on the lattice $\Z^n$.
If the number of reflections $n+1$ is even, their product will be a
translation, invariant under the action of $\G_n$. It gives an invariant
one-dimensional subspace in $\Z^n$. We introduce new coordinates
\begin{align}\label{kx1}
x_i&= k_i- k_{i-1}+ k_{i-2}-\dots +(-1)^{i+1} k_1\\
\label{kx2} k_i&=x_{i-1}+x_i,
\end{align}
so that the last coordinate $x_n$ remains invariant under all $\sigma_i$
if $n+1$ is even.

\begin{lem}
The antisymmetric form
\begin{equation}
J(x,y)=\sum_{i=1}^{n-1}\left( x_iy_{i+1}-x_{i+1}y_i\right)
\end{equation}
is non-degenerate and invariant under the action of $\G_n$
\begin{equation}
J(\sigma x,\sigma y)=J(x,y)=-J(y,x)\,,
\end{equation}
if $n$ is even. It is preserved by $\sigma_i\,,\,i>1$ and has one-dimensional
kernel when $n$ is odd.
\end{lem}
\begin{proof}
The action of the generators $\sigma_i$ of the group $\G_n$ in the coordinates
$x_i$ is
\begin{align}\label{actx1}
\sigma_1&:\left\{\begin{array}{rl}x_{2i}&\mapsto x_{2i}+x_1\\
x_{2i-1}&\mapsto x_{2i-1}\end{array}\right.\\
\label{actx2}
\sigma_i&:\left\{\begin{array}{rl}x_{i-1}&\mapsto x_{i-2}+x_{i-1}-x_i\\
x_j&\mapsto x_j\,j\ne i-1\end{array}\right.\,,\,i\ne 1
\end{align}
so
\begin{multline}
J(\sigma_j x,\sigma_j y)=\sum_{i=1}^{n-1}\left( x_iy_{i+1}-x_{i+1}y_i\right)
+(x_{j-2}-x_j)y_j-(x_{j-2}-x_j)y_{j-2}\\
+x_{j-2}(y_{j-2}-y_j)-x_j(y_{j-2}-y_j)
=J(x,y)
\end{multline}
\begin{multline}
J(\sigma_1 x,\sigma_1 y)=\sum_{i=1}^{\lfloor\frac{n}{2}\rfloor}\bigl(
x_{2i-1}(y_{2i}+y_1)-(x_{2i}+x_1)y_{2i-1}\bigr)\\
+\sum_{i=1}^{\lfloor\frac{n-1}{2}\rfloor}\bigl(
(x_{2i}+x_1)y_{2i+1}-x_{2i+1}(y_{2i}+y_1)\bigr)=\\
J(x,y)+\sum_{i=1}^{\lfloor\frac{n}{2}\rfloor}(x_{2i-1}y_1-x_{2i}y_{2i-1})+
\sum_{i=1}^{\lfloor\frac{n-1}{2}\rfloor}(x_1y_{2i+1}-y_1x_{2i+1})=\\
J(x,y)+\left\{\begin{array}{rl}x_1y_n-x_ny_1&\mbox{if $n$ is odd}\\
0&\mbox{otherwise}\end{array}\right.
\end{multline}

Let $X_1,X_2,\dotsc,X_n$ be the basis vectors in~$\Z^n$, for the coordinates
$x_1,\dotsc,x_n$.
\begin{equation}
\langle X_i,X_j\rangle_J=\delta_{i+1,j}-\delta_{i-1,j}
\end{equation}
The vectors
\begin{equation}
Q_i=X_{2i-1}+X_{2i-3}+\dotsb +X_1\qquad P_i=X_{2i}
\end{equation}
are linearly independent and satisfy
\begin{equation}
\langle Q_i,Q_j\rangle_J=\langle P_i,P_j\rangle_J=0\quad
\langle Q_i,P_j\rangle_J=\delta_{i,j}
\end{equation}
hence they form a canonical basis for the form $J$. If $n$ is odd
$Q_n$ belongs to the kernel of $J$.
\end{proof}

It follows that our group is contained in the symplectic group
over integers $\G_n\subset Sp(2s,\Z)$, if $n=2s$. It is rather unexpected
to appear symplectic group when we consider sets of orthogonal reflections.
It demonstrates an interesting duality between ordered sets of reflections and
ordered sets of transvections. The last are linear transformations with
nontrivial Jordan form, fixing pointwise hyperplanes of codimension one. If
there is a preserved alternating form, the relative position of $n$
transvections $t_1,\dotsc,t_n$ is determined by the matrix
\begin{equation}
H_{ij}=\langle v_i|v_j\rangle,\quad t_i=\Id-\langle v_i|.
\end{equation}
Here $\langle\cdot|\cdot\rangle$ is the alternating form and
$\langle v_i|v_i\rangle=0$. If we write the matrix $H$ as the difference
between an upper triangular matrix $S$ with units on the diagonal and
its transposed, we see that to any set of symplectic transvections there is a
corresponding set of reflections whose Gram matrix is $G=S+S^t$.
The orthogonal reflections with Gram matrix of rank~2 and finite orbit of the
braid group action generate finite dihedral groups, as we have seen. The
braid action on them linearizes and the images of canonical generators of the
braid group are exactly the set of transvections, corresponding to the standard
generators of the Coxeter group $A_n$.

\begin{lem}
The quantity $\delta$ defined by
\begin{equation}
\alpha=\#\{i\,,\quad\frac{ k_i}{\gamma}=1\mod 2\}\,,\qquad
\gamma=\gcd( k_1, k_2,\dotsc, k_{2s})
\end{equation}
\begin{equation}
\delta=|2\alpha-2s-1|\,.
\end{equation}
is an invariant of the action of $\G_{2s}$.
\end{lem}
\begin{proof}
First let us note that the greatest common divisor is an invariant for all
invertible linear transformations with integer coefficients. We may
assume $\gamma=1$ so $\alpha$ is the number of odd $ k_i$-s. If we denote
by $\beta=2s-\alpha$ the number of even coordinates $ k_i$ the action
\eqref{act} preserves $\alpha,\beta$ except for $\sigma_1$ which for odd $ k_1$
converts
\begin{equation}
(\alpha,\beta)\mapsto(\beta+1,\alpha-1)\,.
\end{equation}
This transformation has period 2 and preserves $\delta=|\alpha-\beta-1|$.
\end{proof}

It follows that $\G_{2s}$ is a proper subgroup of $Sp(2n,\Z)$ as the last
preserves only the greatest common divisor $\gamma$ which follows from the
determination of its generators in \cite{Sieg} .
We will prove that the pair $\delta,\gamma$ characterizes
completely the orbits of $\G_{2s}$ in $\Z^{2n}$. For this purpose we first
introduce the natural coordinates $q_i,p_j$ for the form $J$ so that
$Q_i,P_j$ be the basis vectors.
\begin{equation}
\left|\begin{array}{rl}p_i&=x_{2i}\\
q_i&=x_{2i-1}-x_{2i+1}\end{array}\right.\quad
\left|\begin{array}{rl}x_{2i}&=p_i\\
x_{2i-1}&=\sum_{j\ge i}q_i\,.\end{array}\right.
\end{equation}
In these coordinates the invariant form $J$ is given by
\begin{equation}
J(v^{(1)},v^{(2)})=\sum_{i=1}^s q_i^{(1)}p_i^{(2)}-p_i^{(1)}q_i^{(2)}
\end{equation}
and the action of $\G_{2s}$ by
\begin{gather}
\sigma_1:\left\{\begin{array}{rl}q_i&\mapsto q_i\\
p_i&\mapsto p_i+q_1+q_2+\dotsb+q_s\end{array}\right.
\quad\sigma_{2i+1}:\left\{\begin{array}{rl}q_j&\mapsto q_j\\
p_i&\mapsto p_i+q_i\\
p_j&\mapsto p_j\,\quad j\ne i\end{array}\right.\\[5pt]
\sigma_{2i}:\left\{\begin{array}{rl}q_{i-1}&\mapsto q_{i-1}-p_{i-1}+p_i\\
q_i&\mapsto q_i+p_{i-1}-p_i\\
q_j&\mapsto q_j\quad j\ne i,i-1\\
p_j&\mapsto p_j\,.\end{array}\right.
\end{gather}

\begin{thm} There is one-to-one correspondence between the orbits of the
action of $\G_{2s}$ on $\Z^{2s}$ and the pairs $\delta,\gamma$.
The values which these invariants may take are:
\begin{equation}
\gamma\in\Z_+\quad \delta\in\{1,3,5,\dotsc,2s-1\}
\end{equation}
\end{thm}

\begin{proof}
As we already proved $\delta,\gamma$ are preserved by $\G_{2s}$. It remains to
show that $\G_{2s}$ acts transitively on the vectors in $\Z^{2s}$ with equal
quantities $\delta,\gamma$. We will prove it by giving an algorithm
transforming any vector in $\Z^{2s}$ to a canonical representative in each
orbit.

\begin{description}
\item [Step 1.] The pair of transformations $\sigma_2,\sigma_3$ generate
the group $SL(2,\Z)$ on the coordinates $q_1,p_1$. It is
essentially the Euclidean algorithm to transform any vector in this subspace
to $p_1=0$ in the following way. If $|q_1|\ge|p_1|$ the application
of $\sigma_2$ or $\sigma_2^{-1}$ depending whether $q_1p_1>0$ or $q_1p_1<0$
yields $|q'_1|<|q_1|$. If $|q_1|\ge|p_1|$ the application of $\sigma_3^{-1}$
or $\sigma_3$ whenever $q_1p_1>0$ or $q_1p_1<0$ yields $|p'_1|<|p_1|$. The
process of decreasing the absolute values of $q_1,p_1$ finishes when one of
them becomes zero, the other being the greatest common divisor of the initial
$q_1,p_1$. If $q_1=0,p_1\ne 0$ the following transformation gives the desired
result
\begin{equation}
\sigma_2\sigma_3\sigma_2:q_1\mapsto -p_1\,\quad p_1\mapsto q_1
\end{equation}
\item [Step 2.] The pair $\sigma_4,\sigma_5$ acts analogously on the
coordinates $q_2,p_2$, preserving $p_1=0$. Applying the same procedure as the
previous step we can make $p_2=p_1=0$.
\item [Step 3.] Continuing in the same fashion consecutively for all $i<s$
with $\sigma_{2i},\sigma_{2i+1}$; we achieve $p_i=p_{i-1}=\dotsb=p_1=0$.
\item [Step 4.] The pair of transformations $\sigma_{2s},
\sigma_1\sigma_2\dotsm\sigma_{2s-1}\sigma_{2s}\sigma_{2s-1}\dotsm\sigma_1$
act in the same way on the pair $q_s,p_s$, keeping $p_i=0,\,i<s$. Applying
again the Euclidean algorithm we obtain $p_s=0$.
\item [Step 5.] We consider transformations in $\G_{2s}$, preserving
$p_i\equiv 0$.
This subgroup is generated by $U_i=(\sigma_{2i}\sigma_{2i+1}\sigma_{2i})^2$,
$V_i=(\sigma_{2i-1}\sigma_{2i}\sigma_{2i-1})^2$, and
$Z_i=\sigma_{2i+1}\sigma_{2i}
\sigma_{2i+2}\sigma_{2i+1}$ as we will see. In order to make more transparent
the action of these transformations we change the coordinates
\begin{equation}
\left|\begin{array}{rl}\tilde q_i&=\sum\limits_{j\ge i }q_j\\[.5ex]
\tilde p_i&=p_i-p_{i-1}\end{array}\right.\quad
\left|\begin{array}{rl}q_i&=\tilde q_i-\tilde q_{i+1}\\[.5ex]
p_i&=\sum\limits_{j\le i}\tilde p_j\end{array}\right.
\end{equation}
so that the new coordinates $(\tilde q_i,\tilde p_i)$ are again canonical for
$J$, and the subspace $p_i\equiv 0$ coincides with $\tilde p_i\equiv 0$.
In these coordinates
\begin{equation}
V_1:\left\{\begin{array}{rlrl}
\tilde q_1&\mapsto -\tilde q_1\quad&\tilde p_1&\mapsto -\tilde p_1\\
\tilde q_i&\mapsto\tilde q_i\quad&\tilde p_i&\mapsto\tilde p_i\,,\quad i\ne1
\end{array}\right.
\end{equation}
\begin{equation}
U_1:\left\{\begin{array}{rlrlrl}
\tilde q_1&\mapsto -\tilde q_1+2\tilde q_2\quad&\tilde p_1&\mapsto -\tilde p_1
\quad&\tilde p_2\mapsto 2\tilde p_1+\tilde p_2\\
\tilde q_i&\mapsto\tilde q_i\,,\,i\ne1\quad&\tilde p_i&\mapsto\tilde p_i\,,\,
i>2
\end{array}\right.
\end{equation}
\begin{equation}
Z_i:\left\{\begin{array}{rlrlrl}\tilde q_i&\mapsto \tilde q_{i+1}-\tilde p_i
\quad&\tilde q_{i+1}&\mapsto\tilde q_i-\tilde p_{i+1}\quad&
\tilde p_i\leftrightarrow \tilde p_{i+1}\\
\tilde q_j&\mapsto\tilde q_j\quad&
\tilde p_j&\mapsto\tilde p_j\quad&j\ne i,i+1
\end{array}\right.
\end{equation}
As $\tilde p_i\equiv 0$, $Z_i$
transposes $\tilde q_i$ and $\tilde q_{i+1}$. The transformations
$V_1,U_1,Z_1$ generate the principal congruence subgroup of index~2 in
$GL(2,\Z)$  with respect to the action on the 
coordinate pair $\tilde q_1,\tilde q_2$. Using a modified Euclidean
algorithm we may transform to $\tilde q_2=0,\tilde q_1\ge0$ or to
$\tilde q_1=\tilde q_2>0$ in the following way. Change the sign of $\tilde q_1$
or $\tilde q_2$ by $V_1$ or $Z_1V_1Z_1$ to make them both non-negative. Next
apply $Z_1$ if necessary to make $\tilde q_1>\tilde q_2$; and
$V_1U_1^{-1}:\tilde q_1\mapsto\tilde q_1-2\tilde q_2$ decreasing the absolute
value of $\tilde q_1$. Continue in
the same fashion until either $\tilde q_1=\tilde q_2$ or $\tilde q_2=0$.

Repeating the procedure with
the remaining $\tilde q_i$-s as we have all permutations eventually we arrive
at $\tilde q_1=\tilde q_2=\dotsb=\tilde q_l=\gamma,\,
\tilde q_{l+1}=\dotsb \tilde q_s=0$, $\tilde p_1=\dotsb=\tilde p_{2s}=0$.
These will be our canonical representatives in the orbits of the braid group.
In terms of the initial coordinates these are the vectors
$k_1=k_2=\dotsb =k_{2l-2}=0$,
$k_{2l-1}=k_{2l}=\dotsb =k_{2s}=\gamma$ and their invariant $\delta$ is
\begin{equation}
\delta=|4l-2s-1|\,.
\end{equation}
To assure that there are not distinct canonical representatives with equal
quantities $\delta,\gamma$ we check
\begin{equation}
|4l_1-2s-1|=|4l_2-2s-1|,\ l_1,l_2>0 \Rightarrow l_1=l_2
\end{equation}

\end{description}
\end{proof}

\vskip2ex
Next we consider the case of even number of reflections $n=2s+1$.
In this case $x_n$ is an invariant, measuring the length of the
translation  $r_1r_2\dotsm r_{n+1}$.

Although in this case the
antisymmetric form $J$ is not preserved by $\sigma_1$ we will use its natural
coordinates $q_i,p_i$ introduced by \eqref{kx1}--\eqref{kx2}.

\begin{lem}
The triple $\gamma,\delta,x_{2s+1}$, defined by
\begin{equation}
\alpha=\#\{i,\quad\frac{ k_i}{\gamma}=1\mod 2\}\,,\qquad
\gamma=\gcd( k_1, k_2,\dotsc, k_{2s+1})
\end{equation}
\begin{equation}
\delta=|2\alpha-2s-2|\quad x_{2s+1}= k_1- k_2+\dotsc+ k_{2s+1}
\end{equation}
remains invariant under the action of $\G_{2s+1}$
on $\Z^{2s+1}$.
\end{lem}

The proof is the same as with the even-dimensional case. The invariance of
$x_{2s+1}$ was mentioned when these coordinates were introduced \eqref{kx1}.

\begin{thm}
There is one-to-one correspondence between the orbits of $\G_{2s+1}$ in
$\Z^{2s+1}$ and the triples $\gamma,\delta,x_{2s+1}$. The values which these
invariants may take are
\begin{equation}
\gamma\in\Z_+\,,\quad \delta\in\{1,3,\dotsc,2n-1\}\,,\quad x_{2s+1}=s\gamma\,,
\,s\in\N
\end{equation}
\end{thm}
\begin{proof}
First let us note that we may invert the signs of all $\phi_i$ to assure
$x_{2s+1}\ge 0$.
Again it remains to prove only the transitivity of the action of $\G_{2s+1}$ on
vectors with equal $\gamma,\delta,x_{2s+1}$. We will use the same algorithm
for transformation to a canonical representative in each orbit with the
following changes:
\begin{description}
\item [Step 4.] The pair of transformations $\sigma_{2s},\sigma_{2s+1}$ should
be used in order to make $p_s=0$.
\item [Step 5.] The change of coordinates should not involve the invariant
$q_{s+1}=x_{2s+1}$ so
\begin{equation}
\left|\begin{array}{rl}\tilde q_i&=\sum\limits_{i\le j\le s}q_j\\[.5ex]
\tilde p_i&=p_i-p_{i-1}\end{array}\right.\quad
\left|\begin{array}{rl}q_i&=\tilde q_i-\tilde q_{i+1}\,,\quad x\ne s\\[.5ex]
p_i&=\sum\limits_{j\le i}\tilde p_j\end{array}\right.\,.
\end{equation}
In these coordinates
\begin{equation}
U_s:\left\{\begin{array}{rlrl}
\tilde q_s&\mapsto -\tilde q_s\quad&\tilde p_s&\mapsto -\tilde p_s\\
\tilde q_i&\mapsto\tilde q_i\quad&\tilde p_i&\mapsto\tilde p_i\,,\quad i\ne s
\end{array}\right.
\end{equation}
\begin{equation}
U_1:\left\{\begin{array}{rlrlrl}
\tilde q_1&\mapsto -\tilde q_1+2\tilde q_2\quad&\tilde p_1&\mapsto -\tilde p_1
\quad&\tilde p_2\mapsto -2\tilde p_1+\tilde p_2\\
\tilde q_i&\mapsto\tilde q_i\,,\,i\ne1\quad&\tilde p_i&\mapsto\tilde p_i\,,\,
i>2
\end{array}\right.
\end{equation}
\begin{equation}
Z_i:\left\{\begin{array}{rlrlrl}\tilde q_i&\mapsto \tilde q_{i+1}-\tilde p_i
\quad&\tilde q_{i+1}&\mapsto\tilde q_i-\tilde p_{i+1}\quad&
\tilde p_i\leftrightarrow \tilde p_{i+1}\\
\tilde q_j&\mapsto\tilde q_j\quad&
\tilde p_j&\mapsto\tilde p_j\quad&j\ne i,i+1
\end{array}\right.
\end{equation}
Using repeatedly the modified Euclidean algorithm we may transform each
vector to $\tilde q_1=\tilde q_2=\dotsb=\tilde q_l=\gamma',\,
\tilde q_{l+1}=\dotsb \tilde q_s=0$. Notice that up to now, there were used
only transformations not involving $\sigma_1$, which are exactly those,
preserving the form $J$.
\item [Step 6.]
If $\gamma'$ is a divisor of $q_{s+1}$ the procedure stops.
Otherwise we apply
\begin{equation}
V=\sigma_1\sigma_2\dotsm\sigma_{2s}\sigma_{2s+1}\sigma_{2s}^{-1}\dotsm
\sigma_1^{-1}:\left\{\begin{array}{rl}\tilde q_i&\mapsto
\tilde q_i+\tilde q_{s+1}-\sum\tilde p_j\\
\tilde p_i&\mapsto\tilde p_i\end{array}\right.\,.
\end{equation}
By definition $\gcd(\gamma',q_{s+1})=\gamma$ which implies
$\gcd(\gamma'+q_{s+1},q_{s+1})=\gamma$. If $l<s$ the repeating of {\bf Step 5}
will end the procedure. If $l=n$ the application of $V^r$ will transform
$\gamma'\mapsto\gamma'+rq_{s+1}$. We may take the minimal non-negative value
of $\gamma'$ for the canonical representative in this orbit.
\end{description}

Expressing
\begin{equation}
\delta=|2\alpha-2s-2|=|4l-2s-2|
\end{equation}
we check whether for different $l_1\ne l_2$ may correspond the same
invariant $\delta$:
\begin{equation}
|4l_1+2a-2s-2|=|4l_2+2a-2s-2|\,,\quad 1\le l\le s
\quad\Rightarrow l_2=l_1,s+1-a-l_1.
\end{equation}
Not only $\delta$ but also $\alpha$ remains constant during
Steps 1--5 i.e. when $\sigma_1$ doesn't act. As
$l_1,l_2$ correspond to $\alpha_1,\alpha_2=\beta_1+1$ and the action of
$\sigma_1$ on the canonical element changes $\alpha_1\mapsto\alpha_2$,
it is clear, that the vectors with $l_1\ne l_2,\,\delta_1=\delta_2$
belong to the same orbit of $\G_{2s+1}$.
\end{proof}

We return to the modular case
\begin{equation}\label{mod}
G_{ij}=2\cos\left(\frac{k_i-k_j}{m}\gamma\pi\right)\,,\quad k_i\in\Z_m,\,
0\le\gamma<m,\,\gcd(\gamma,m)=1\,.
\end{equation}
The invariant $\delta$ may be defined only if $m\in2\Z$. It presented an
obstruction in the algorithm for the canonical representative in each orbit
at the fifth step, when we obtained a congruence subgroup of index 2 in
$SL(2,\Z)$, generated by $U_1,V_1,Z_1$, whose action on $\tilde q_1,
\tilde q_2$ is given by the matrices
\begin{equation}
V_1=\begin{pmatrix}-1&0\\0&1\end{pmatrix}\quad
Z_1=\begin{pmatrix}0&1\\1&0\end{pmatrix}\quad
U_1V_1^{-1}=\begin{pmatrix}1&2\\0&1\end{pmatrix}
\end{equation}
If $m$ is odd certain power of the last matrix will be
$\bigl(\begin{smallmatrix}1&1\\0&1\end{smallmatrix}\bigr)\mod m$ so we have the
full $SL(2,\Z_m)$.

\begin{cor}
Every finite orbit of $B_{n+1}$ on the rank 2, $(n+1)\times(n+1)$ Gram
matrices \eqref{mod} is in one-to-one correspondence with
the following quantities
\begin{description}
\item $m$, $2\le m$;
\item $\gamma$, $1\le\gamma<m$, $\gcd(\gamma,m)=1$;
\item $\delta\in\{1,3,\dotsc,\left\lfloor\frac{n}{2}\right\rfloor-1\}$ if
$m\in2\Z$;
\item $x$, $0\le x<m$ if $n\not\in2\Z$.
\end{description}
\end{cor}

Now we are able to describe the nature of the group $\G_n$. First let us note
the well-known fact that the group of permutations of $n+1$ elements $S_{n+1}$
can be represented by $n+1\times n+1$ matrices
\begin{equation}
M_{ij}=\delta_{i\pi(j)}\,,\quad \pi\in S_{n+1}\,.
\end{equation}
If we consider binary matrices over the field $\Z_2$ 
the permutations of $n+1$ elements can also be represented by $n\times n$
matrices
\begin{equation}
M_{ij}=\delta_{i\pi(j)}+\delta_{0\pi(j)}\,,\quad S_{n+1}\ni\pi:
\{0,1,2,\dotsc,n\}\to\{0,1,2,\dotsc,n\}\,,
\end{equation}
which are non-degenerate and closed under multiplication
\begin{multline}
(M_1\cdot M_2)_{ij}=\sum_{k=1}^n(\delta_{i\pi_1(k)}+\delta_{0\pi_1(k)})
(\delta_{k\pi_2(j)}+\delta_{0\pi_2(j)})\\
=(1-\delta_{0\pi_2(j)})\delta_{i\pi_1\pi_2(j)}+
(1-\delta_{i\pi_1(0)})\delta_{0\pi_2(j)}\\
+(1-\delta_{0\pi_2(j)})\delta_{0\pi_1\pi_2(j)}+
(1-\delta_{0\pi_1(0)})\delta_{0\pi_2(j)}\\
=\delta_{i\pi_1\pi_2(j)}+\delta_{0\pi_1\pi_2(j)}+
2\delta_{0\pi_2(j)}(1-\delta_{i\pi_1(0)}-\delta_{0\pi_1(0)})\\
\equiv \delta_{i\pi_1\pi_2(j)}+\delta_{0\pi_1\pi_2(j)} \mod 2
\end{multline}

Matrices of the transformations \eqref{act} acting on $\Z_2^n$, correspond
to the generating
transpositions $(0,1),(1,2),\dotsc,(n-1,n)$ in the above representation of
$S_{n+1}$. This demonstrates an isomorphism between the matrices of $\G_n$
over $\Z_2$ and the group $S_{n+1}$. We have seen that
$\G_{2s}\subset Sp(2s,\Z)$. Moreover $\G_{2s}(\Z_{2k})\subset Sp(2n,\Z_{2k})$ and
$\G_{2s}(\Z_{2k+1})=Sp(2n,\Z_{2k+1})$. The transformation
\begin{equation}
\tau:\left\{\begin{array}{lll}
q_i&\mapsto&q_{i+1}\,,\quad i<n\\
q_n&\mapsto&q_1\\
p_i&\mapsto&p_{i+1}\,,\quad i<n\\
p_n&\mapsto&p_1\\ \end{array}\right.
\end{equation}
intermixes the orbits with different $\delta$ and belongs to $Sp(2s,\Z)$
therefore when added to $\G_{2s}$ will generate the whole $Sp(2s,\Z)$
(see \cite{Sieg}).

In order to find the number of co-classes of $\G_{2s}$ in $Sp(2s,\Z)$ it is
enough to divide the number of elements in the last group over the field $\Z_2$
by that of the first:
\begin{equation}
Sp(2s,\Z)/\G_{2s}=\frac{2^{s^2}\prod_{k=1}^s(2^{2k}-1)}{(2s+1)!}\,.
\end{equation}

The group $\G_{2s+1}$ preserve the coordinate $x_{2s+1}$ and in the
corresponding basis its elements have the form
\begin{equation}
g=\begin{pmatrix}g_0&g_1\\0&1\end{pmatrix}\,,\quad g_0\in Sp(2s,\Z)\,,
\quad g_1\in\Z^{2s}
\end{equation}
The index of $\G_{2s+1}$ in $Sp(2s,\Z)\ltimes\Z^{2s}$ is
\begin{equation}
Sp(2s,\Z)\ltimes\Z^{2s}/\G_{2s+1}=
\frac{2^{s^2+2s}\prod_{k=1}^s(2^{2k}-1)}{(2s+2)!}\,.
\end{equation}


\begin{thebibliography}{99}
\bibitem{Arn} Arnol'd, V.,{\it Remark on the branching of hyperelliptic
integrals as functions of the parameters}, Functional Anal. Appl.
{\bf 2}(1968), 187--189
\bibitem{Art1} Artin, E., {\it Theorie der Z\"opfe}, Abh. Math. Sem. Hamburg,
Vol. 4, 1926, pp.47--72
\bibitem{Art2} Artin, E., {\it Theory of braids}, Annals of Math., Vol. 48,
1946, pp. 101--126
\bibitem{Bir} Birman J.S., {\it Braids, links, and mapping class groups},
Princeton, N.J., Princeton University Press, 1974
\bibitem{Bol} Bolibruch, A. A. {\it On isomonodromic deformations of
Fuchsian systems}. J. Dynam. Control Systems  3  (1997),  no. 4, 589--604
\bibitem{Bo} Bourbaki N., Lie Groups and Lie Algebras: Chapters 4-6,
Addison-Wesley; 1st edition (1975)
\bibitem{Bur} Burau, W. ,{\it\"Uber Zopfgruppen und gleichsinnig verdrilte
Verkettungen}, Abh. Math. Sem. Hanischen Univ. 11(1936) 171--178
\bibitem{D1} B.Dubrovin, {\it Geometry of 2D topological field theories},
in: Integrable Systems and Quantum Groups, Montecatini, Terme, 1993.
Editors: M.Francaviglia, S. Greco. Springer Lecture Notes in
Math. {\bf 1620} (1996), 120--348.
\bibitem{D2} B. Dubrovin, {\it Painlev\'e transcendents in two-dimensional
topological field theory.} In: ``The Painlev\'e property: 100 years later'', 287--412,
CRM Ser. Math. Phys., Springer, New York, 1999.
\bibitem{taniguchi} B. Dubrovin, 
{\it Flat pencils of metrics and Frobenius manifolds}, math.DG/9803106,
In: Proceedings of 1997 Taniguchi Symposium ``Integrable Systems
and Algebraic Geometry", Editors M.-H. Saito, Y.Shimizu and K.Ueno,
47-72. World Scientific, 1998.
\bibitem{icm} B. Dubrovin, {\it Geometry and analytic theory of
Frobenius manifolds},
%\newline
math/9807034, Proceedings of ICM98, Vol. 2, 315-326. 
\bibitem{duality}B.Dubrovin, {\it On almost duality for Frobenius manifolds},
math.DG/0307374; to appear in AMS Transl.
\bibitem{DM} Dubrovin B., Mazzocco M., {\it Monodromy of certain Painlev\'e-VI
transcendents and reflection groups}, Invent. Math.  141  (2000),  no. 1,
55--147.
\newline
math.AG/9806056
\bibitem{Hum} J. E. Humphreys, {\it Reflection Groups and Coxeter Groups},
Cambridge University Press; Reprint edition (1993)
\bibitem{Hum1}J. E. Humphreys, {\it Arithmetic groups},
Lecture Notes in Mathematics, 789, Springer, Berlin, (1980).
\bibitem{Cox} Coxeter H.S.M., {\it Regular Polytopes}, Dover Pubns;
3rd edition (1973)
\bibitem{Ince}Ince E.L., {\it Ordinary Differential Equations}, London
- New York etc., Longmans, Green and Co., 1927.
\bibitem{Sieg}Siegel C.L., {\it Topics in complex function theory}, Vol.2
\bibitem{JMU} Jimbo, M., Miwa, T. and Ueno, K., {\it Monodromy preserving
deformation of linear ordinary differential equations with rational
coefficients I}, Physica 2D (1981), 306-352.
\bibitem{Sch} Schlesinger, L.,
{\it\"Uber eine Klasse von Differentialsystemmen beliebliger Ordnung
mit festen kritischer Punkten},
J. F\"ur Math., {\bf 141},(1912), pp 96--145
\bibitem{U1} K.Ueno , {\it Monodromy preserving deformation of
linear differential equations with irregular singular points}.  
Proc. Japan Acad. Ser. A Math. Sci.  56  (1980), no. 3, 97--102.


\end{thebibliography}
\end{document}